\documentclass[sigconf]{acmart} 
\pdfoutput=1
\acmConference[ICSE 2024]{46th International Conference on Software Engineering}{April 2024}{Lisbon, Portugal}

\setcopyright{acmlicensed}
\copyrightyear{2024}
\acmYear{2024}
\acmDOI{XXXXXXX.XXXXXXX}

\acmISBN{978-1-4503-XXXX-X/18/06}

\usepackage[T1]{fontenc}
\usepackage[utf8]{inputenc}
\usepackage{graphicx}
\usepackage{textcomp}
\usepackage{xspace}
\usepackage{pgfplots}
\pgfplotsset{compat=1.18}
\usepackage{tikz}
\usepackage{rotating}
\usepackage{multirow}
\usepackage{booktabs}
\usepackage{subcaption}
\usepackage{paralist}
\usepackage{adjustbox}
\usepackage{amsthm}
\usepackage{siunitx}
\newcommand{\toolname}[1]{\textsc{#1}\xspace}
\newcommand{\utbotpyhton}{\toolname{UTBotPython}}

\newcommand{\pynguin}{\toolname{Pynguin}}
\newcommand{\ghostwriter}{\toolname{Hypothesis Ghostwriter}}
\newcommand{\klara}{\toolname{Klara}}

\newcommand{\tensorflow}{TensorFlow\xspace}
\newcommand{\django}{Django\xspace}
\newcommand{\flask}{Flask\xspace}
\newcommand{\numpy}{Numpy\xspace}
\newcommand{\scikit}{scikit-learn\xspace}
\newcommand{\ansible}{Ansible\xspace}
\newcommand{\spark}{Spark\xspace}

\newcommand{\ie}{i.e.,\xspace}

\makeatletter
\DeclareRobustCommand{\iscircle}{\mathord{\mathpalette\is@circle\relax}}
\newcommand\is@circle[2]{%
  \begingroup
  \sbox\z@{\raisebox{\depth}{$\m@th#1\bigcirc$}}%
  \sbox\tw@{$#1\square$}%
  \resizebox{!}{\ht\tw@}{\usebox{\z@}}%
  \endgroup
}
\makeatother

\begin{document}

\title{SBFT Tool Competition 2024 - Python Test Case Generation Track}

\author{Nicolas Erni}
\affiliation{
  \institution{Zurich University of Applied Sciences}
  \city{}
  \country{Switzerland}}

\author{Al-Ameen Mohammed Ali Mohammed}
\affiliation{
  \institution{Zurich University of Applied Sciences}
  \city{}
  \country{Switzerland}}

\author{Christian Birchler}
\affiliation{
  \institution{Zurich University of Applied Sciences \\ University of Bern}
  \city{}
  \country{Switzerland}}
  
\author{Pouria Derakhshanfar}
\affiliation{
  \institution{JetBrains}
  \city{}
  \country{The Netherlands}}

\author{Stephan Lukasczyk  }
\affiliation{
  \institution{University of Passau}
  \city{}
  \country{Germany}}

\author{Sebastiano Panichella }
\affiliation{
  \institution{Zurich University of Applied Sciences}
  \city{}
  \country{Switzerland}}

\renewcommand{\shortauthors}{Erni et al.}


\begin{abstract} 
Test case generation (TCG) for Python poses distinctive challenges due to the language's dynamic nature and the absence of strict type information.
Previous research has successfully explored automated unit TCG for Python, with solutions outperforming random test generation methods.
Nevertheless, fundamental issues persist, hindering the practical adoption of existing test case generators.
To address these challenges, we report on the organization, challenges, and results of the first edition of the Python Testing Competition.
Four tools, namely \utbotpyhton, \klara, \ghostwriter, and \pynguin were executed on a benchmark set consisting of 35~Python source files sampled from 7~open-source Python projects for a time budget of 400~seconds. 
We considered one configuration of each tool for each test subject and evaluated the tools' effectiveness in terms of code and mutation coverage.
This paper describes our methodology, the analysis of the results together with the competing tools, and the challenges faced while running the competition experiments.
\end{abstract}


\maketitle

\section{Introduction}

This year, we organized the first edition of the Python SBFT Tool Competition.
The competition has the goal to experiment with testing tools for a diversified set of systems and domains.
We invited researchers to participate in the competition with their unit test generation tools for Python~\cite{pythonic-idioms}.
The tools are evaluated against benchmarks concerning code and mutation coverage as similarly done in previous SBFT tool competitions for Java~\cite{Devroey2022}.



\section{The Python Testing Competition}
\label{sec:introduction-Python}
The first edition of the Python Testing Tool Competition received one submitted tool: \utbotpyhton~\cite{utbot-python}.
Furthermore, similarly to previous/contemporary testing competition editions of SBFT~\cite{DBLP:conf/sbst/PanichellaGZR21,JahangirovaT23,SBFT-UAV2024}, we used various different baseline tools, namely \pynguin~\cite{Lukasczyk_2022}, \ghostwriter~\cite{hypothesis}, and \klara~\cite{klara} for comparison.
Each test generation tool has been executed on (\ie generated test cases for) 35 Python source code files sampled from 7 open-source projects on GitHub, which are \tensorflow~\cite{tensorflow}, \django~\cite{django}, \flask~\cite{flask}, \numpy~\cite{numpy}, \scikit~\cite{scikit-learn}, \ansible~\cite{ansible}, and \spark~\cite{spark}.
Eventually, with the submitted tool and the three baseline tools evaluated on seven benchmark projects, we get a broad performance overview of the state-of-the-art test generation tools for Python.

To guarantee a fair comparison among the competing tools, the execution of the tools for generating test suites and their evaluation has been carried out by using an infrastructure hosted on GitHub\footnote{\url{https://github.com/ThunderKey/python-tool-competition-2024}} and Zenodo~\cite{competition-results-zenodo}.
Each tool implements the same interface provided by the aforementioned infrastructure code.
The competing tools have been compared by using a set of various coverage metrics, such as line, branch, and mutant coverage.
For the comparison, all tools got a one-time budget of 400 seconds to generate test cases for the aforementioned benchmark source files.


\subsection{Benchmark subjects}
\label{sec:subjects-Python}

Similarly to previous editions of SBFT tool competitions~\cite{DBLP:conf/sbst/PanichellaGZR21,JahangirovaT23}, the selection of the projects and Python files under test to use as benchmark for test case generation has been done by considering three criteria:
(i) projects must belong to different application domains; (ii) projects must be open-source for replicability purposes; and (iii) files must not have increasing complexity. 

We focused on popular open-source projects on GitHub written in Python.
Specifically, we picked: 
\begin{itemize}
    \item \tensorflow\footnote{\url{https://github.com/tensorflow/tensorflow}}: an end-to-end open source platform for machine learning;
    \item \django\footnote{\url{https://github.com/django/django}}: a high-level Python web framework to build complex data-driven websites;
    \item \flask\footnote{\url{https://github.com/pallets/flask}}: a lightweight WSGI web application framework;
    \item \numpy\footnote{\url{https://github.com/numpy/numpy}}: a package for scientific computing with Python;
    \item \scikit\footnote{\url{https://github.com/scikit-learn/scikit-learn}}: is a Python module for machine learning;
    \item \ansible\footnote{\url{https://github.com/ansible/ansible}}: an IT automation system for configuration management, application development, and cloud provisioning;
    \item \spark\footnote{\url{https://github.com/apache/spark}}: an analytics engine for large-scale data processing.
\end{itemize}

Based on the time and resources available for running the competition, we have only sampled a limited number of files.
Specifically for the third selection criteria, we gathered a dataset of Python files, ensuring each file contained at least one function and maintained an average code length of 20 lines.
Furthermore, our selection criteria explicitly excluded files that possess the potential to terminate processes, notably, those leveraging the \texttt{os}\footnote{\url{https://docs.python.org/3/library/os.html}} module, as well as files linked to libraries with significant security implications, particularly those with capabilities to alter the file system through write or delete operations.


\begin{table}[t]
    \caption{Description of the benchmarks.}
    \label{tab:overview}
    \centering
    \small
    \begin{tabular}{lSS} \toprule
    \textbf{Project} & \textbf{\# Python Files} & \textbf{\# Sampled Python Files} \\ \midrule
    \tensorflow & 3135& 5\\
    \django  &2771&5\\
    \flask  & 82&5\\
    \numpy  & 581&5\\
    \scikit  & 926& 5\\
    \ansible  & 1559& 5\\
    \spark  & 1134& 5\\ \midrule
    \textbf{\textsc{Total}} & 10188& 35\\ \bottomrule
    \end{tabular}
\end{table}

\subsection{Competing tools}
\label{sec:tools-Python}

Four tools are competing in the first edition: \utbotpyhton~\cite{utbot-python}, \pynguin~\cite{Lukasczyk_2022}, \ghostwriter~\cite{hypothesis}, and \klara~\cite{klara}.
We received one submitted tool, \utbotpyhton, and use \ghostwriter, \klara, and \pynguin as baseline tools.
All four tools implement different approaches to generate test cases:

\begin{itemize}

\item \utbotpyhton~\cite{utbot-python} generates test cases based on precise code analysis using a symbolic execution engine paired with a smart fuzzing technique.

\item \pynguin~\cite{Lukasczyk_2022} is a framework that implements several search-based unit test generation algorithms.
For the competition it uses its implementation of the DynaMOSA~\cite{Panichella2018} algorithm, which used to emit test suites with the highest coverage values compared to the other algorithms \pynguin provides~\cite{LukasczykKF23}.

\item \toolname{Hypothesis}~\cite{hypothesis} is a library for creating property-based tests. Its \toolname{Ghostwriter} module provides a way to automatically generating these property-based testing features.

\item \klara~\cite{klara} is a set of static analysis tools to automatically generate test cases, based on AST analysis.

\end{itemize}

\subsection{Methodology}
\label{sec:methodology-Python}

The methodology followed to run the competition is similar to the one adopted in the previous editions of the SBFT tool competitions for Java~\cite{DBLP:conf/sbst/PanichellaGZR21,JahangirovaT23}.
It is important to remark that, due to time and resource constraints and the number of competing tools (four considering also the baseline approaches), we only considered a one-time budget of 400 seconds.

\textbf{Public contest repository.}
The complete competition infrastructure is released under a GPL-3.0 license and is available on GitHub.\footnote{\url{https://github.com/ThunderKey/python-tool-competition-2024}}
Specifically, the repository contains the set of Python files contributing to the first edition and the detailed summary of the results obtained by running each tool for each time budget. 
The competition participants are required to implement a given interface given by the infrastructure code.
When implemented in the provided interface, the infrastructure code can evaluate in a straightforward manner the test generators based on a set of coverage criteria, which are described in more detail below.
 
\textbf{Test generation and time budget.}
Each tool has been executed four times against each benchmark target file to account for the randomness of the test case generation process~\cite{10.1002/stvr.1486}.
All executions got a time budget of 400 seconds to ensure a fair and feasible comparison with the available execution environment.

\textbf{Execution environment.}
The infrastructure performed a total of 560~executions, \ie
\(35\text{ Python files} \times 4\text{ tools} \times 1\text{ time budget} \times 4\text{ repetitions}\), to use for statistical analysis.
To ensure a fair comparison, we ran each tool on the same dedicated machine, \ie a virtual machine instance equipped with four vCPUs, \SI{7.8}{\giga\byte} of RAM and \SI{155}{\giga\byte} of memory running Ubuntu 22.04.3 LTS. 
For all the competing tools, we were able to run the planned number of executions.

\textbf{Metrics computation.}
We compared the performance of the competing tools based on line, branch, and mutation coverage metrics.
Specifically, to compute both line and branch coverage metrics, we relied on \textsc{PyTest}~\cite{pytest} and its \textsc{pytest-cov}\footnote{\url{https://pytest-cov.readthedocs.io/en/latest/}} plugin,
an open-source framework for writing unit tests.
For mutation analysis, we relied on \textsc{MutPy}~\cite{mutpy} and \textsc{Cosmic Ray}~\cite{cosmic-ray}, considering five minutes as the maximum amount of time available for mutation analysis for each Python file.


\subsection{Competition's results and analysis}
\label{sec:results-Python}

\begin{table}[t]
    \centering
    \caption{Statistics on the number of generated test cases for each tool and the time budget of 400\,s after four runs.}
    \label{tab:tests}
    \small
    \resizebox{\linewidth}{!}{
    \begin{tabular}{l S S S S} \toprule
    \textbf{Tool} & \multicolumn{4}{c}{\textbf{\# Test Cases}} \\
                  & \textbf{min} & \textbf{mean} & \textbf{median} & \textbf{max} \\ \midrule
    \utbotpyhton  & 2            & 3             & 3               & 4            \\ 
    \pynguin      & 1            & 4.67          & 3               & 10           \\
    \ghostwriter  & 1            & 5.33          & 3               & 12           \\
    \klara        & 1            & 2.33          & 3               & 3            \\ \bottomrule
    \end{tabular}}
\end{table}

\begin{figure*}
    \centering
    \begin{tikzpicture}
  \begin{axis}[
  ybar,
  width=18cm,
  height=5cm,
  bar width=10pt,
  enlargelimits=0.1,
  enlarge y limits=0.2,
  legend style={at={(0.55, 0.99), font=\tiny},
  anchor=north,legend columns=2},
  ylabel={\scriptsize Line coverage},
  symbolic x coords={
  \tensorflow,
  \django,
  \flask,
  \numpy,
  \scikit,
  \ansible,
  \spark
  },
  xtick=data,
  x tick label style={rotate=0, anchor=north, font=\scriptsize},
  nodes near coords,
  nodes near coords align={vertical},
  nodes near coords style={
      anchor=west,
      rotate=90,
      font=\scriptsize,
      #1
  }
  ]

  \addplot coordinates {
  (\tensorflow,0.182)
  (\django,0)
  (\flask,0)
  (\numpy,0.2135)
  (\scikit,0)
  (\ansible,0)
  (\spark,0)
  };
  
  \addplot coordinates {
  (\tensorflow,0.365818)
  (\django,0.319)
  (\flask,0)
  (\numpy,0.16)
  (\scikit,0)
  (\ansible,0.49)
  (\spark,0)
  };
  
  \addplot coordinates {
  (\tensorflow,0.1636)
  (\django,0.2825)
  (\flask,0)
  (\numpy,0.2336)
  (\scikit,0)
  (\ansible,0.3421)
  (\spark,0)
  };
  
  \addplot coordinates {
  (\tensorflow,0)
  (\django,0.16)
  (\flask,0.0526)
  (\numpy,0)
  (\scikit,0)
  (\ansible,0)
  (\spark,0)
  };
  \legend{\utbotpyhton,\pynguin,\ghostwriter,\klara}
  
  \end{axis}
  \end{tikzpicture}
  \begin{tikzpicture}
  \begin{axis}[
  ybar,
  width=18cm,
  height=5cm,
  bar width=10pt,
  enlargelimits=0.1,
  enlarge y limits=0.2,
  legend style={at={(0.55, 0.99), font=\tiny},
  anchor=north,legend columns=2},
  ylabel={\scriptsize Branch converage},
  symbolic x coords={
  \tensorflow,
  \django,
  \flask,
  \numpy,
  \scikit,
  \ansible,
  \spark
  },
  xtick=data,
  x tick label style={rotate=0, anchor=north, font=\scriptsize},
  nodes near coords,
  nodes near coords align={vertical},
  nodes near coords style={
      anchor=west,
      rotate=90,
      font=\scriptsize,
      #1
  }
  ]

  \addplot coordinates {
  (\tensorflow,0.6)
  (\django,0.4)
  (\flask,0.5)
  (\numpy,0.0833)
  (\scikit,0.2)
  (\ansible,0.4)
  (\spark,0)
  };
  
  \addplot coordinates {
  (\tensorflow,0.78)
  (\django,0.7)
  (\flask,0.75)
  (\numpy,0.083332)
  (\scikit,0.2)
  (\ansible,0.84375)
  (\spark,0.0)
  };
  
  \addplot coordinates {
  (\tensorflow,0.6)
  (\django,0.4)
  (\flask,0.5)
  (\numpy,0.112492)
  (\scikit,0.2)
  (\ansible,0.61875)
  (\spark,0.0)
  };
  
  \addplot coordinates {
  (\tensorflow,0.6)
  (\django,0.4)
  (\flask,0.5)
  (\numpy,0.0)
  (\scikit,0.2)
  (\ansible,0.4)
  (\spark,0.0)
  };
  \legend{\utbotpyhton,\pynguin,\ghostwriter,\klara}
  
  \end{axis}
  \end{tikzpicture}
  

  \begin{tikzpicture}
  \begin{axis}[
  ybar,
  width=18cm,
  height=5cm,
  bar width=10pt,
  enlargelimits=0.1,
  enlarge y limits=0.2,
  legend style={at={(0.55, 0.99), font=\tiny},
  anchor=north,legend columns=2},
  ylabel={\scriptsize Mutation score},
  symbolic x coords={
  \tensorflow,
  \django,
  \flask,
  \numpy,
  \scikit,
  \ansible,
  \spark
  },
  xtick=data,
  x tick label style={rotate=0, anchor=north, font=\scriptsize},
  nodes near coords,
  nodes near coords align={vertical},
  nodes near coords style={
      anchor=west,
      rotate=90,
      font=\scriptsize,
      #1
  }
  ]

  \addplot coordinates {
  (\tensorflow,0.6)
  (\django,0.2)
  (\flask,0.5)
  (\numpy,0.0466)
  (\scikit,0)
  (\ansible,0.4)
  (\spark,0)
  };
  
  \addplot coordinates {
  (\tensorflow,0.6)
  (\django,0.2)
  (\flask,0.5)
  (\numpy,0.04)
  (\scikit,0.0)
  (\ansible,0.5030)
  (\spark,0.0)
  };
  
  \addplot coordinates {
  (\tensorflow,0.6)
  (\django,0.2)
  (\flask,0.5)
  (\numpy,0.0)
  (\scikit,0.0)
  (\ansible,0.4)
  (\spark,0.0)
  };
  
  \addplot coordinates {
  (\tensorflow,0.6)
  (\django,0.2)
  (\flask,0.5)
  (\numpy,0.0)
  (\scikit,0.0)
  (\ansible,0.4)
  (\spark,0.0)
  };
  \legend{\utbotpyhton,\pynguin,\ghostwriter,\klara}
  
  \end{axis}
  \end{tikzpicture}

  
    \caption{Average coverage and mutation score for each test generation tool after 4 runs.}
    \label{fig:barplot-benchmarks}
\end{figure*}
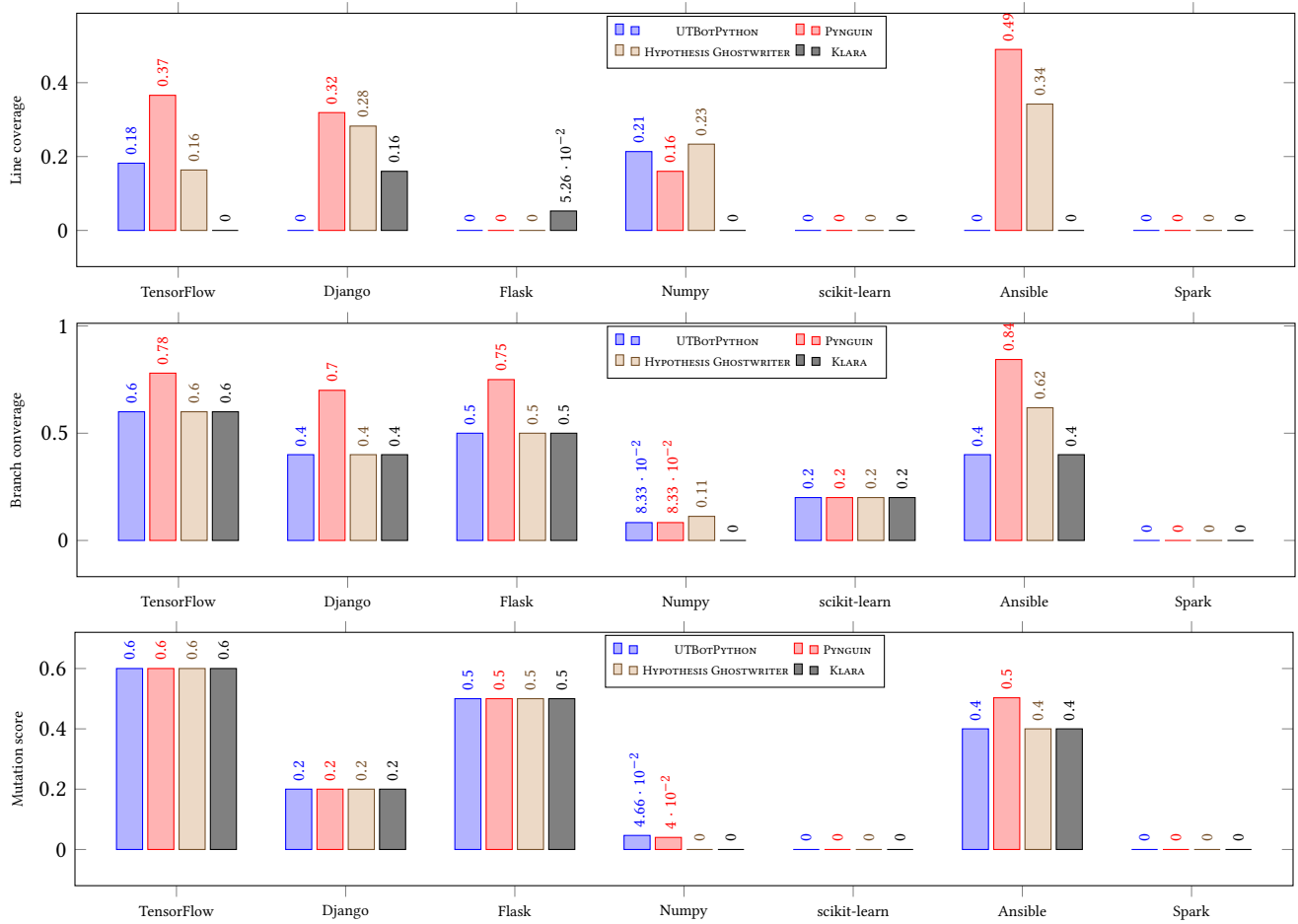

\begin{table}[t]
    \centering
    \caption{Average performance scores of the tools among all benchmarks.}
    \label{tab:scores}
    \small
    \resizebox{\linewidth}{!}{
    \begin{tabular}{l S[round-mode=figures,round-precision=3] S[round-mode=figures,round-precision=3] S[round-mode=figures,round-precision=3] S[round-mode=figures,round-precision=3]} \toprule
      \textbf{Tool} & \multicolumn{3}{c}{\textbf{Coverage (\%)}}          & \textbf{Total Score} \\
                    & \textbf{Line} & \textbf{Branch} & \textbf{Mutation} &                      \\ \midrule
      \pynguin      & 26.69         & 27.40           & 5.2805            & 1.026033             \\
      \utbotpyhton  & 10.10         & 6.63            & 6.6007            & 0.497597             \\
      \ghostwriter  & 17.99         & 9.04            & 2.7228            & 0.469544             \\
      \klara        & 3.45          & 2.41            & 2.2277            & 0.171784             \\ \bottomrule
    \end{tabular}
    }
\end{table}

Each tool generates a different number of test cases.
We executed each tool four times with a time budget of \SI{400}{\second}.
Numbers are rounded to three significant digits, if appropriate.
Table \ref{tab:tests} presents for each tool the minimum, mean, median, and maximum number of the generated test cases.
We observed that \ghostwriter generates, on average, most test cases followed by \pynguin, \utbotpyhton, and then \klara.

We computed different metrics, such as the line, branch coverage, and mutation score to evaluate the quality of the generated test cases.
Table~\ref{tab:scores} reports the average percentage of lines, branches, and mutants being covered by the tools after four executions.
Regarding line coverage, \pynguin has the highest value with \SI{26.7}{\percent}.
Meanwhile, \klara had the lowest with \SI{3.45}{\percent}.
In terms of branch coverage, \pynguin also has the highest score with more than \SI{27}{\percent}, while \klara has again the lowest with \SI{2.41}{\percent}.
Furthermore, Table~\ref{tab:scores} also reports the average mutation coverage, which is the ratio between the number of mutants that were killed by at least one test and the total number of mutants being generated. 
Eventually, we report the final scores of the tools based on a time budget of \SI{400}{\second}.
The formula~\cite{Devroey2022} for the final score has been created and improved during the previous editions of the tool competition and takes into account the line and branch coverage and the mutation score used by the generator.
In terms of ranking, we have \pynguin as first, followed by \utbotpyhton, \ghostwriter, and \klara.

In addition to the general performance evaluation, we also evaluated the tools' performance for each benchmark project individually.
Figure~\ref{fig:barplot-benchmarks} depicts the various performance metrics of each tool among all benchmarks.
We observed that the tools could not create any tests for certain benchmarks.
For example, no tool was able to generate tests for the \spark project.
We can see this due to the fact that all metrics were zero for this specific benchmark.
Interestingly, for the \scikit project, we still got some branch coverage but no line coverage.
This observation contradicts the general perception that no branch can be covered when no line is covered.
We think this branch coverage might occur due to empty \texttt{if} clauses where the condition is true but no executable statement is in the block except for the \texttt{pass} keyword.
Furthermore, in the case of the \spark benchmark, only \klara could cover a small number of lines.

A potential cause for this low line coverage might still be the code complexity of those aforementioned benchmark projects although it was already a selection criteria not having an increasing complexity (Section~\ref{sec:subjects-Python}).
We argue for future editions of the competition, the tools should investigate the feasibility to handle complex file structures from real benchmark projects.
Furthermore, it could be interesting to investigate to what extent certain newly introduced Python language features affect the tools in its performances.

\section{Conclusions and Final Remarks of the Python Testing Tool Competition}
\label{sec:conclusion-Python}
This year marks the first edition of the Python Unit Testing Competition.
We received one tool, namely \utbotpyhton, which competes against three baseline tools, namely, \pynguin, \ghostwriter, and \klara.
As per results of this year, the best-performing tool overall is \pynguin followed by \utbotpyhton, and \ghostwriter while \klara seems to perform the worst on the selected benchmark subject files.
The analysis of collected results by the organizers of the competition and by the participants revealed that for some of the generated test suites, it was not possible to generate tests or compute coverage and mutation analysis. 
The most likely cause of this is the files are too complex, contain many relative imports, or lack modularity.
We plan to investigate this issue further to identify the definite root cause of the problem and to perform a fix for the next editions of the competition in order to provide better criteria for the right type of files.
In addition, we envision several other possibilities for improvement, such as extending the list of criteria used for the evaluation, such as performance awareness~\cite{8865437} and readability~\cite{PanichellaPBZG16,RaniPLSN21} scores. 

\section{Data availability}
We provide all code and detailed results on Zenodo~\cite{competition-results-zenodo}.

\section*{Acknowledgments}
We thank the participants 
of the competitions for their invaluable
contribution. 
We thank the Horizon 2020 (EU Commission) and Innosuisse support for the projects COSMOS (DevOps for Complex Cyber-physical Systems, Project No. 957254-COSMOS) and ARIES (Project 45548.1 IP-ICT).

\balance
\bibliographystyle{ACM-Reference-Format}
\bibliography{references}


\begin{thebibliography}{26}


\ifx \showCODEN    \undefined \def \showCODEN     #1{\unskip}     \fi
\ifx \showDOI      \undefined \def \showDOI       #1{#1}\fi
\ifx \showISBNx    \undefined \def \showISBNx     #1{\unskip}     \fi
\ifx \showISBNxiii \undefined \def \showISBNxiii  #1{\unskip}     \fi
\ifx \showISSN     \undefined \def \showISSN      #1{\unskip}     \fi
\ifx \showLCCN     \undefined \def \showLCCN      #1{\unskip}     \fi
\ifx \shownote     \undefined \def \shownote      #1{#1}          \fi
\ifx \showarticletitle \undefined \def \showarticletitle #1{#1}   \fi
\ifx \showURL      \undefined \def \showURL       {\relax}        \fi
\providecommand\bibfield[2]{#2}
\providecommand\bibinfo[2]{#2}
\providecommand\natexlab[1]{#1}
\providecommand\showeprint[2][]{arXiv:#2}

\bibitem[ans(2024)]%
        {ansible}
 \bibinfo{year}{2024}\natexlab{}.
\newblock \bibinfo{booktitle}{\emph{Ansible}}.
\newblock
\urldef\tempurl%
\url{https://github.com/ansible/ansible}
\showURL{%
\tempurl}


\bibitem[cos(2024)]%
        {cosmic-ray}
 \bibinfo{year}{2024}\natexlab{}.
\newblock \bibinfo{booktitle}{\emph{cosmic-ray}}.
\newblock
\urldef\tempurl%
\url{https://cosmic-ray.readthedocs.io/en/latest/}
\showURL{%
\tempurl}


\bibitem[dja(2024)]%
        {django}
 \bibinfo{year}{2024}\natexlab{}.
\newblock \bibinfo{booktitle}{\emph{Django}}.
\newblock
\urldef\tempurl%
\url{https://github.com/django/django}
\showURL{%
\tempurl}


\bibitem[fla(2024)]%
        {flask}
 \bibinfo{year}{2024}\natexlab{}.
\newblock \bibinfo{booktitle}{\emph{flask}}.
\newblock
\urldef\tempurl%
\url{https://github.com/pallets/flask}
\showURL{%
\tempurl}


\bibitem[kla(2024)]%
        {klara}
 \bibinfo{year}{2024}\natexlab{}.
\newblock \bibinfo{booktitle}{\emph{Klara}}.
\newblock
\urldef\tempurl%
\url{https://github.com/usagitoneko97/klara}
\showURL{%
\tempurl}


\bibitem[num(2024)]%
        {numpy}
 \bibinfo{year}{2024}\natexlab{}.
\newblock \bibinfo{booktitle}{\emph{Numpy}}.
\newblock
\urldef\tempurl%
\url{https://github.com/numpy/numpy}
\showURL{%
\tempurl}


\bibitem[pyt(2024)]%
        {pytest}
 \bibinfo{year}{2024}\natexlab{}.
\newblock \bibinfo{booktitle}{\emph{pytest}}.
\newblock
\urldef\tempurl%
\url{https://docs.pytest.org/en}
\showURL{%
\tempurl}


\bibitem[sci(2024)]%
        {scikit-learn}
 \bibinfo{year}{2024}\natexlab{}.
\newblock \bibinfo{booktitle}{\emph{scikit-learn}}.
\newblock
\urldef\tempurl%
\url{https://github.com/scikit-learn/scikit-learn}
\showURL{%
\tempurl}


\bibitem[spa(2024)]%
        {spark}
 \bibinfo{year}{2024}\natexlab{}.
\newblock \bibinfo{booktitle}{\emph{Spark}}.
\newblock
\urldef\tempurl%
\url{https://github.com/apache/spark}
\showURL{%
\tempurl}


\bibitem[ten(2024)]%
        {tensorflow}
 \bibinfo{year}{2024}\natexlab{}.
\newblock \bibinfo{booktitle}{\emph{TensorFlow}}.
\newblock
\urldef\tempurl%
\url{https://github.com/tensorflow/tensorflow}
\showURL{%
\tempurl}


\bibitem[utb(2024)]%
        {utbot-python}
 \bibinfo{year}{2024}\natexlab{}.
\newblock \bibinfo{booktitle}{\emph{UTBotPython}}.
\newblock
\urldef\tempurl%
\url{https://github.com/UnitTestBot/UTBotPythonSBFT2024}
\showURL{%
\tempurl}


\bibitem[Alexandru et~al\mbox{.}(2018)]%
        {pythonic-idioms}
\bibfield{author}{\bibinfo{person}{Carol~V. Alexandru}, \bibinfo{person}{Jos\'{e}~J. Merchante}, \bibinfo{person}{Sebastiano Panichella}, \bibinfo{person}{Sebastian Proksch}, \bibinfo{person}{Harald~C. Gall}, {and} \bibinfo{person}{Gregorio Robles}.} \bibinfo{year}{2018}\natexlab{}.
\newblock \showarticletitle{On the Usage of Pythonic Idioms}. In \bibinfo{booktitle}{\emph{ACM SIGPLAN International Symposium on New Ideas, New Paradigms, and Reflections on Programming and Software}}. \bibinfo{publisher}{{ACM}}, \bibinfo{pages}{1–11}.
\newblock
\urldef\tempurl%
\url{https://doi.org/10.1145/3276954.3276960}
\showDOI{\tempurl}


\bibitem[Arcuri and Briand(2014)]%
        {10.1002/stvr.1486}
\bibfield{author}{\bibinfo{person}{Andrea Arcuri} {and} \bibinfo{person}{Lionel Briand}.} \bibinfo{year}{2014}\natexlab{}.
\newblock \showarticletitle{A Hitchhiker’s Guide to Statistical Tests for Assessing Randomized Algorithms in Software Engineering}.
\newblock \bibinfo{journal}{\emph{Software Testing, Verification \& Reliability}} \bibinfo{volume}{24}, \bibinfo{number}{3} (\bibinfo{year}{2014}), \bibinfo{pages}{219--250}.
\newblock
\showISSN{0960-0833}
\urldef\tempurl%
\url{https://doi.org/10.1002/stvr.1486}
\showDOI{\tempurl}


\bibitem[Derezinska and Halas(2014)]%
        {mutpy}
\bibfield{author}{\bibinfo{person}{Anna Derezinska} {and} \bibinfo{person}{Konrad Halas}.} \bibinfo{year}{2014}\natexlab{}.
\newblock \showarticletitle{Experimental Evaluation of Mutation Testing Approaches to Python Programs}. In \bibinfo{booktitle}{\emph{International Conference on Software Testing, Verification and Validation}}. \bibinfo{publisher}{{IEEE} Computer Society}, \bibinfo{pages}{156--164}.
\newblock
\urldef\tempurl%
\url{https://doi.org/10.1109/ICSTW.2014.24}
\showDOI{\tempurl}


\bibitem[Devroey et~al\mbox{.}(2023)]%
        {Devroey2022}
\bibfield{author}{\bibinfo{person}{Xavier Devroey}, \bibinfo{person}{Alessio Gambi}, \bibinfo{person}{Juan~Pablo Galeotti}, \bibinfo{person}{Ren{\'{e}} Just}, \bibinfo{person}{Fitsum~Meshesha Kifetew}, \bibinfo{person}{Annibale Panichella}, {and} \bibinfo{person}{Sebastiano Panichella}.} \bibinfo{year}{2023}\natexlab{}.
\newblock \showarticletitle{{JUGE:} An infrastructure for benchmarking Java unit test generators}.
\newblock \bibinfo{journal}{\emph{Software Testing, Verification and Reliability}} \bibinfo{volume}{33}, \bibinfo{number}{3} (\bibinfo{year}{2023}).
\newblock
\urldef\tempurl%
\url{https://doi.org/10.1002/STVR.1838}
\showDOI{\tempurl}


\bibitem[Erni et~al\mbox{.}(2024)]%
        {competition-results-zenodo}
\bibfield{author}{\bibinfo{person}{Nicolas Erni}, \bibinfo{person}{Al-Ameen~Mohammed Ali~Mohammed}, \bibinfo{person}{Christian Birchler}, \bibinfo{person}{Pouria Derakhshanfar}, \bibinfo{person}{Stephan Lukasczyk}, {and} \bibinfo{person}{Sebastiano Panichella}.} \bibinfo{year}{2024}\natexlab{}.
\newblock \bibinfo{title}{SBFT Tool Competition 2024 - Python Test Case Generation Track}.
\newblock
\newblock
\urldef\tempurl%
\url{https://doi.org/10.5281/zenodo.10554259}
\showDOI{\tempurl}


\bibitem[Grano et~al\mbox{.}(2021)]%
        {8865437}
\bibfield{author}{\bibinfo{person}{Giovanni Grano}, \bibinfo{person}{Christoph Laaber}, \bibinfo{person}{Annibale Panichella}, {and} \bibinfo{person}{Sebastiano Panichella}.} \bibinfo{year}{2021}\natexlab{}.
\newblock \showarticletitle{Testing with Fewer Resources: An Adaptive Approach to Performance-Aware Test Case Generation}.
\newblock \bibinfo{journal}{\emph{IEEE Transactions on Software Engineering}} \bibinfo{volume}{47}, \bibinfo{number}{11} (\bibinfo{year}{2021}), \bibinfo{pages}{2332--2347}.
\newblock


\bibitem[Jahangirova and Terragni(2023)]%
        {JahangirovaT23}
\bibfield{author}{\bibinfo{person}{Gunel Jahangirova} {and} \bibinfo{person}{Valerio Terragni}.} \bibinfo{year}{2023}\natexlab{}.
\newblock \showarticletitle{{SBFT} Tool Competition 2023 - Java Test Case Generation Track}. In \bibinfo{booktitle}{\emph{International Workshop on Search-Based and Fuzz Testing}}. \bibinfo{publisher}{{IEEE}}, \bibinfo{pages}{61--64}.
\newblock
\urldef\tempurl%
\url{https://doi.org/10.1109/SBFT59156.2023.00025}
\showDOI{\tempurl}


\bibitem[Khatiri et~al\mbox{.}(2024)]%
        {SBFT-UAV2024}
\bibfield{author}{\bibinfo{person}{Sajad Khatiri}, \bibinfo{person}{Prasun Saurabh}, \bibinfo{person}{Timothy Zimmermann}, \bibinfo{person}{Charith Munasinghe}, \bibinfo{person}{Christian Birchler}, {and} \bibinfo{person}{Sebastiano Panichella}.} \bibinfo{year}{2024}\natexlab{}.
\newblock \showarticletitle{{SBFT} Tool Competition 2024 - CPS-UAV Test Case Generation Track}. In \bibinfo{booktitle}{\emph{International Workshop on Search-Based and Fuzz Testing}}. \bibinfo{publisher}{{ACM}}.
\newblock


\bibitem[Lukasczyk and Fraser(2022)]%
        {Lukasczyk_2022}
\bibfield{author}{\bibinfo{person}{Stephan Lukasczyk} {and} \bibinfo{person}{Gordon Fraser}.} \bibinfo{year}{2022}\natexlab{}.
\newblock \showarticletitle{Pynguin: automated unit test generation for Python}. In \bibinfo{booktitle}{\emph{International Conference on Software Engineering: Companion}}. \bibinfo{publisher}{{ACM}}, \bibinfo{pages}{168--172}.
\newblock
\urldef\tempurl%
\url{https://doi.org/10.1145/3510454.3516829}
\showDOI{\tempurl}


\bibitem[Lukasczyk et~al\mbox{.}(2023)]%
        {LukasczykKF23}
\bibfield{author}{\bibinfo{person}{Stephan Lukasczyk}, \bibinfo{person}{Florian Kroi{\ss}}, {and} \bibinfo{person}{Gordon Fraser}.} \bibinfo{year}{2023}\natexlab{}.
\newblock \showarticletitle{An empirical study of automated unit test generation for Python}.
\newblock \bibinfo{journal}{\emph{Empirical Software Engineering}} \bibinfo{volume}{28}, \bibinfo{number}{2} (\bibinfo{year}{2023}), \bibinfo{pages}{36}.
\newblock
\urldef\tempurl%
\url{https://doi.org/10.1007/S10664-022-10248-W}
\showDOI{\tempurl}


\bibitem[Maciver and Hatfield{-}Dodds(2019)]%
        {hypothesis}
\bibfield{author}{\bibinfo{person}{David Maciver} {and} \bibinfo{person}{Zac Hatfield{-}Dodds}.} \bibinfo{year}{2019}\natexlab{}.
\newblock \showarticletitle{Hypothesis: {A} new approach to property-based testing}.
\newblock \bibinfo{journal}{\emph{Journal of Open Source Software}} \bibinfo{volume}{4}, \bibinfo{number}{43} (\bibinfo{year}{2019}), \bibinfo{pages}{1891}.
\newblock
\urldef\tempurl%
\url{https://doi.org/10.21105/JOSS.01891}
\showDOI{\tempurl}


\bibitem[Panichella et~al\mbox{.}(2018)]%
        {Panichella2018}
\bibfield{author}{\bibinfo{person}{Annibale Panichella}, \bibinfo{person}{Fitsum~Meshesha Kifetew}, {and} \bibinfo{person}{Paolo Tonella}.} \bibinfo{year}{2018}\natexlab{}.
\newblock \showarticletitle{Automated Test Case Generation as a Many-Objective Optimisation Problem with Dynamic Selection of the Targets}.
\newblock \bibinfo{journal}{\emph{IEEE Transactions on Software Engineering}} \bibinfo{volume}{44}, \bibinfo{number}{2} (\bibinfo{year}{2018}), \bibinfo{pages}{122--158}.
\newblock
\urldef\tempurl%
\url{https://doi.org/10.1109/TSE.2017.2663435}
\showDOI{\tempurl}


\bibitem[Panichella et~al\mbox{.}(2021)]%
        {DBLP:conf/sbst/PanichellaGZR21}
\bibfield{author}{\bibinfo{person}{Sebastiano Panichella}, \bibinfo{person}{Alessio Gambi}, \bibinfo{person}{Fiorella Zampetti}, {and} \bibinfo{person}{Vincenzo Riccio}.} \bibinfo{year}{2021}\natexlab{}.
\newblock \showarticletitle{{SBST} Tool Competition 2021}. In \bibinfo{booktitle}{\emph{International Workshop on Search-Based Software Testing}}. \bibinfo{publisher}{{IEEE}}, \bibinfo{pages}{20--27}.
\newblock
\urldef\tempurl%
\url{https://doi.org/10.1109/SBST52555.2021.00011}
\showDOI{\tempurl}


\bibitem[Panichella et~al\mbox{.}(2016)]%
        {PanichellaPBZG16}
\bibfield{author}{\bibinfo{person}{Sebastiano Panichella}, \bibinfo{person}{Annibale Panichella}, \bibinfo{person}{Moritz Beller}, \bibinfo{person}{Andy Zaidman}, {and} \bibinfo{person}{Harald~C. Gall}.} \bibinfo{year}{2016}\natexlab{}.
\newblock \showarticletitle{The impact of test case summaries on bug fixing performance: an empirical investigation}. In \bibinfo{booktitle}{\emph{International Conference on Software Engineering}}. \bibinfo{publisher}{{ACM}}, \bibinfo{pages}{547--558}.
\newblock
\urldef\tempurl%
\url{https://doi.org/10.1145/2884781.2884847}
\showDOI{\tempurl}


\bibitem[Rani et~al\mbox{.}(2021)]%
        {RaniPLSN21}
\bibfield{author}{\bibinfo{person}{Pooja Rani}, \bibinfo{person}{Sebastiano Panichella}, \bibinfo{person}{Manuel Leuenberger}, \bibinfo{person}{Andrea~Di Sorbo}, {and} \bibinfo{person}{Oscar Nierstrasz}.} \bibinfo{year}{2021}\natexlab{}.
\newblock \showarticletitle{How to identify class comment types? {A} multi-language approach for class comment classification}.
\newblock \bibinfo{journal}{\emph{Journal of Systems and Software}}  \bibinfo{volume}{181} (\bibinfo{year}{2021}), \bibinfo{pages}{111047}.
\newblock
\urldef\tempurl%
\url{https://doi.org/10.1016/J.JSS.2021.111047}
\showDOI{\tempurl}


\end{thebibliography}

\end{document}